\documentclass[12pt,a4paper,twoside]{article}
\usepackage[latin1]{inputenc}
\usepackage[english]{babel}
\usepackage[margin=3cm]{geometry}

\usepackage{hyperref}
\usepackage{graphicx}
\usepackage{times}
\usepackage{fancyhdr}
\usepackage{epsfig}
\usepackage{multirow}
\usepackage{amssymb,amsfonts,amsmath}
\usepackage{enumerate}
\usepackage{cite}

\newcommand{\ntag}[1]{{\sc #1}}
\newcommand{\panino}[1]{\left\langle #1\right\rangle}

\usepackage[hang,small,bf]{caption}

\begin{document}

\renewcommand{\multirowsetup}{\centering}

\thispagestyle{empty} 

\begin{center}
 {\Large\bfseries\sffamily
 Infection propagator approach to compute epidemic thresholds on temporal networks: impact of immunity and of limited temporal resolution\\
 }
 \vspace{1cm}\large
{Eugenio Valdano\textsuperscript{1},
Chiara Poletto\textsuperscript{1},\\
and Vittoria Colizza\textsuperscript{1,2}}\\
\vspace{1.1cm}
{\footnotesize
\textsuperscript{1}Sorbonne Universit\'es, UPMC Univ Paris 06, INSERM, Institut Pierre Louis d'\'Epid\'emiologie et de Sant\'e Publique (IPLESP~UMRS 1136), F75012, Paris, France\\
\textsuperscript{2}ISI Foundation, Torino, Italy.\\
}
\end{center}

 \vspace{2.2cm}

{\sffamily\footnotesize
\centering{\bfseries Abstract}\\
The epidemic threshold of a spreading process indicates the condition for the occurrence of the wide spreading regime, thus representing a predictor of the network vulnerability to the epidemic. Such threshold depends on the natural history of the disease and on the pattern of contacts of the network with its time variation. Based on the theoretical framework introduced in (Valdano et al. PRX 2015) for a susceptible-infectious-susceptible model, we formulate here an infection propagator approach to compute the epidemic threshold accounting for more realistic effects regarding a varying force of infection per contact, the presence of immunity, and a limited time resolution of the temporal network. We apply the approach to two temporal network models and an empirical dataset of school contacts. We find that permanent or temporary immunity do not affect the estimation of the epidemic threshold through the infection propagator approach. Comparisons with numerical results show the good agreement of the analytical predictions. Aggregating the temporal network rapidly deteriorates the predictions, except for slow diseases once the heterogeneity of the links is preserved. Weight-topology correlations are found to be the critical factor to be preserved to improve accuracy in the prediction.
}

\newpage


\section{Introduction}
\label{intro}

The concept of epidemic threshold is fundamental in infectious disease modeling~\cite{Anderson1992,Keeling2008}. When a pathogen is seeded in a population, a critical transmissibility exists below which the spread rapidly ceases. Such a threshold is a combined property of the disease natural history and of the network of interactions along which transmission can occur. In the physics literature such interplay has been typically studied for the family of susceptible-infected-susceptible and susceptible-infected-recovered  models on networks~\cite{PastorSatorras2001,Wang2003,Gomez2010,Cohen2000,Newman2002,Boguna2013,Castellano2010,Goltsev2012,Barrat2008}. Several analytical approaches based, for instance, on the heterogenous mean field approximation~\cite{PastorSatorras2001}, on percolation theory~\cite{Cohen2000,Newman2002} and on Markov processes~\cite{Wang2003,Gomez2010} have been developed to study the transition from early extinction to epidemic. 

Extensive work has been done under the assumption of spreading time scales either much slower or much faster than the one characteristic of the underlying network -- the two regimes called  annealed and quenched, respectively~\cite{PastorSatorras2001,Gomez2010}. In  recent years,  the massive amount of empirical information on networks has showed that such assumption does not hold in many cases~\cite{Bansal2010,Isella2011,Miritello2011,Rocha2010,Bajardi2011,Butts2009,Karsai2011,Holme2012} and that the network dynamics presents features (e.g. memory, bursty activation, heterogeneities in node activity)  affecting the resulting spreading processes~\cite{Isella2011,Miritello2011,Karsai2011,Rocha2013,Ferreri2014,Iribarren2009,Vazquez2007,Eames2004,Perra2012,Volz2009,Gross2006,Taylor2012}. 

The majority of studies addressing so far the impact of network dynamics on the epidemic spread through the analytical calculation
of the epidemic threshold  are all based on synthetic models of the network evolution, valid under context-specific assumptions~\cite{Ferreri2014,Eames2004,Perra2012,Volz2009,Gross2006,Taylor2012,Zhao2010}. To fill this gap, we have introduced in~\cite{Valdano2015} a method to compute the epidemic threshold for a susceptible-infectious-susceptible (SIS) process on a generic discrete-time temporal network, assuming the knowledge of its sequence of adjacency matrices. The approach is rooted in a multi-layer representation~\cite{DeDomenico2013,Kivela2014} of the temporal network that preserves the network causality. It employs a tensor formulation that integrates both spreading and network dynamics and  allows for the analytical solution of the linearized Markov chain description of the spreading process. Such framework extends in this way the quenched approach to the time-varying case, through a multilayer transformation.

The lack of assumption on the network substrate makes such a tool a candidate for assessing the vulnerability to epidemic invasion of real systems  for which time-varying contact data relevant for epidemic transmissions are collected~\cite{Isella2011,Miritello2011,Rocha2010,Bajardi2011,Butts2009,Karsai2011,Holme2012,Vanhems2013,Konschake2013,Fournet2014,Obadia2015}. At the same time, it allows a systematic exploration of the structural and temporal factors characterizing the time-evolving network that are responsible for  sustained spreading. To allow the use of this framework to a variety of different settings and epidemic conditions, we assess here
the applicability of the approach in describing realistic diseases and its robustness with respect to network properties induced by data collection procedures and availability. By considering empirical and synthetic model contact data, we discuss how a varying force of infection along a given link and its direction impact the computation of the threshold for a SIS dynamics. Then, we formulate the
approach for more realistic disease natural histories, considering susceptible-infectious-recovered (SIR) and susceptible-infectious-recovered-susceptible (SIRS) compartmental models. This allows us to account for an additional ingredient~--~immunity following infection, either permanent or temporary~--~representing an important key feature of many diseases. Finally, we address the problem of limited temporal resolution in the knowledge or availability of the network dynamics, for which contacts occurring within a given time interval are aggregated~\cite{Ribeiro2013}. By focusing on an empirical network of time-varying contacts among individuals at school, we quantify the accuracy and reliability of the estimation of the epidemic threshold testing increasing aggregations, to provide quantitative and qualitative information on the specific temporal structures responsible for observed biases. 

\section{Infection propagator approach for a weighted directed temporal network}

We consider a SIS model~\cite{Anderson1992,Keeling2008} where hosts, represented by nodes in the network, can be either in the susceptible or the infectious state. We assume the process to unfold in discrete time on a weighted directed temporal network,  comprising a finite number $T$ of snapshots, each one with a weighted adjacency matrix $\mathbf{W}_t$. The entry $W_{t,ij}$ encodes the weight of the directed link from $i$ to $j$ at time step $t$. At each time step, infectious nodes spontaneously recover with probability $\mu$, returning to the susceptible state. While infectious, nodes can transmit the infection to susceptible neighbors with a probability that depends both on the weight of the link and on the intrinsic transmissibility of the pathogen $\lambda$, representing the probability of transmission when link weight is equal to $1$. We model this by introducing a transmission matrix $\mathbf{\Lambda}_t$, function of both $\lambda$ and $\mathbf{W}_t$, that encodes transmission probabilities. When the network is unweighted ($W_{t,ij}=A_{t,ij}=0,1$), the entries of the matrix $\mathbf{\Lambda}_t$ are simply given by $\Lambda_{t,ij}=\lambda A_{t,ij}$. If the network is weighted, several choices are possible to model transmission along the weighted link. Here we consider a binomial process for the infection
\begin{equation}
\Lambda_{t,ij} = 1-\left(1-\lambda\right)^{W_{t,ij}},
\label{eq:weight_binomial}
\end{equation}
as it is typically assumed, for example, in the spread of livestock infections between premises where the weight represents the number of animals moved between farms~\cite{Bajardi2012}, or in the context of social networks, where weight may encode force-of-infection in terms of either contact frequency or duration~\cite{Machens2013}.

We start from the microscopic Markov chain approach, or {\itshape quenched mean field approach}, developed for static networks~\cite{Wang2003,Gomez2010}. According to this, the equations describing the SIS propagation on a generic static network with $N$ nodes and adjacency matrix $\mathbf{W}$ are
\begin{equation}
  p_{t,i} = 1-\left[ 1- \left(1-\mu\right)p_{t-1,i} \right]  \prod_j \left(1- \Lambda_{ji} p_{t-1,j} \right),
 \label{eq:basic1Gen}
\end{equation}
where $p_{t,i}$ is the probability that node $i$ is in the infectious state at time $t$, and $\mathbf{\Lambda}$ is the static transmission matrix. We remark that Eq.~(\ref{eq:basic1Gen}) relies on the assumption that no dynamical correlations exist among infection probabilities of neighbouring nodes~\cite{Gleeson2012}.

The microscopic Markov chain model of Eq.~(\ref{eq:basic1Gen}) is widely adopted in different fields~\cite{Boguna2013,VanMieghem2011}. For both directed and undirected networks~\cite{VanMieghem2013,Li2013} the study of its asymptotic state yields the derivation of the epidemic threshold in terms of the spectral radius of the transmission matrix $\rho[\mathbf{\Lambda}]$, namely the modulus of the largest eigenvalue of $\mathbf{\Lambda}$. The threshold is the  value of $\lambda$ for which the following holds: $\rho[\mathbf{\Lambda}]=\mu$. In the unweighted case, this is equivalent to the well-known relation $\left(\lambda/\mu \right) = 1/\rho[\mathbf{A}]$, where $\rho[\mathbf{A}]$  is the spectral radius of the adjacency matrix~\cite{Wang2003,Gomez2010}.

In order to extend this approach to temporal networks we need to take into account the time dependence of $\mathbf{\Lambda}$. The Markov chain equations of the process read in this case:
\begin{equation}
 p_{t,i} = 1 - \left[1-\left(1-\mu\right)p_{t-1,i}\right] \prod_j \left(1- \Lambda_{t-1,ji} p_{t-1,j} \right).
 \label{eq:temporalSIS}
\end{equation}
We enforce the existence of the asymptotic solution of the infection process in a generic temporal network by imposing periodic boundary conditions for network dynamics, i.e. $\mathbf{W}_{T+1}\equiv \mathbf{W}_{1}$. Given that $T$ is arbitrary, this causes no loss in generality. We also tested that it would affect the epidemic threshold estimation only for rather small values of $T$, also when complex temporal dynamics are considered~\cite{Valdano2015}. Contrary to the static case, now the asymptotic solutions of Eq.~(\ref{eq:temporalSIS}) are periodic of period $T$. 

\begin{figure}[!ht]
\begin{center}
\resizebox{0.75\textwidth}{!}{\includegraphics{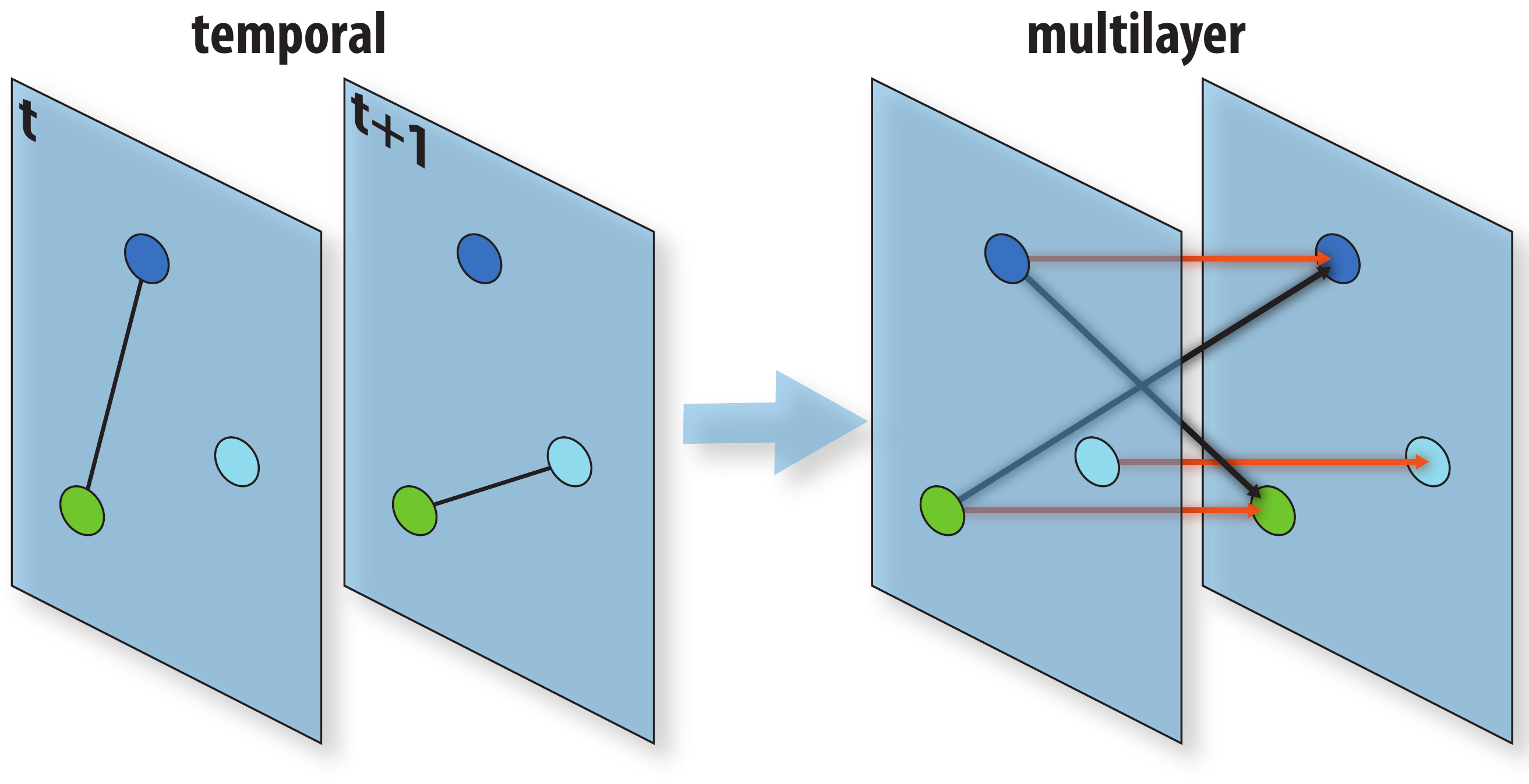}}
\end{center}
\caption{
{\bfseries Multi-layer representation of the temporal network.} We consider a temporal network of $3$ nodes a $3$ time steps (left). This network is mapped onto a multi-layer representation (right). Each node points to its future self (orange arrows), and to the future image of its present neighbors (black arrows).
}
\label{fig:figure_multilayer}
\end{figure}

We develop the formalism introduced in~\cite{Valdano2015}, and compute the epidemic threshold for a generic weighted directed temporal network.
 We define a new representation of the SIS dynamics on a temporal network by employing a multi-layer representation~\cite{DeDomenico2013,Kivela2014,Wehmuth2014}. We map the temporal network to the tensor space $\mathbb{R}^{N}\otimes \mathbb{R}^T$, where each node is identified by the pair of indices $(i,t)$, corresponding  to the node label $i$ and the time frame $t$ respectively. Layer $t$ thus contains the images of the nodes corresponding to time step $t$. The specific multi-layer representation of the temporal network is built according to the following rules~\cite{Valdano2015}: 
 \begin{enumerate}[(i)]
 \item each node, at time $t$, is connected to its future self-image at time $t+1$; 
 \item if $i$ is connected to $j$ at time $t$ with weight $w$, then we connect $i$ at time $t$ to $j$ at time $t+1$, and $j$ at time $t$ to $i$ at time $t+1$, both with weight $w$.
 \end{enumerate}
These rules define a tensor representation of a weighted multilayer network~\cite{Kivela2014,Mucha2010}. We stress that no links connect nodes on the same layer, as contacts in the temporal networks are mapped onto the inter-layer links of the multilayer object (rule (ii)). The resulting network is thus multipartite, since only pairs of nodes belonging to different layers are linked together. Figure~\ref{fig:figure_multilayer} provides a schematic representation of this multilayer mapping.
The adjacency representation of the resulting multilayer network has as entries $\hat{W}_{tt',ij} = \delta_{t,t'+1}\left[  \delta_{ij} + W_{t,ij}  \right]$.
The proposed mapping from the network temporal sequence to a multilayer object provides an \textit{ad hoc} representation of the temporal network that preserves the causality of the temporal network and that it lends itself to the integration of the infection and recovery processes. The transformation for the links (rule (ii)) is similar to the one introduced in~\cite{Scholtes2014}, and is here introduced to model the infection process along a time-stamped link. In addition, we also need to consider the connection between each node and its future self (rule (i)) to model the recovery process of each infected node. We can therefore define the transmission tensor $\mathbf{M}$, whose entries are defined as:
\begin{equation}
 M_{tt',ij} = \delta_{t,t'+1}\left[  \left(1-\mu\right)\delta_{ij} + \Lambda_{t,ij}  \right].
 \label{eq:matr_multipartite}
\end{equation}
 $\mathbf{M}$ contains the  transmission terms $\mathbf{\Lambda}_t$ and the recovery term. $\mathbf{M}$ also introduces a simplified expression for Eq.~(\ref{eq:temporalSIS}). Using the supra-adjacency matrix formalism~\cite{DeDomenico2013,Cozzo2013b,Wang2013}, we can flatten out the multilayer representation using the following mapping: $(i,t) \to \alpha=  Nt+i$, with $\alpha$ running in $\left\{ 1, ... , NT \right\}$, allowing us to write $\mathbf{M}$ in matrix form 
\[
\mathbf{M} = 
\left(
\begin{smallmatrix} 
0 & 1-\mu+\mathbf{\Lambda}_1 & 0 & \cdots & 0 \\
0 & 0 & 1-\mu+\mathbf{\Lambda}_2 & \cdots & 0 \\
\vdots & \vdots & \vdots & \vdots & \vdots \\
0 & 0 & 0 & \cdots & 1-\mu+\mathbf{\Lambda}_{T-1} \\
1-\mu+\mathbf{\Lambda}_{T} & 0 & 0 & \cdots & 0 \\
\end{smallmatrix} 
\right).
\]
$\mathbf{M}$ provides a representation of the topological and temporal dimensions underlying the dynamics of Eq.~(\ref{eq:temporalSIS}), in terms of a $NT\times NT$ transmission matrix that encodes both pathogen transmission and recovery.
The Markov process is now represented in $\mathbb{R}^{NT}$ by the state vector $\hat{p}_\alpha(\tau)$, i.e. the probability of each node to be infectious at each time step $t$ included in a $1$-period long interval, $\left[\tau T, (\tau +1) T \right]$. Consistently, Eq.~(\ref{eq:temporalSIS}) becomes 
\begin{equation}
 \hat{p}_\alpha (\tau) = 1-  \prod_\beta \left[ 1-  M_{\beta \alpha} \hat{p}_\beta \right (\tau-1) ].
 \label{eq:temporal2}
\end{equation}
Given that vector $\hat{p}$ encodes a $1$-period configuration, the $T$-periodic asymptotic state of the SIS process is now mapped into the steady state $\hat{p}_\alpha(\tau)=\hat{p}_\alpha(\tau-1)$. The latter can be recovered as solution of the equilibrium equation:
\begin{equation}
 \hat{p}_\alpha = 1-  \prod_\beta \left( 1-  M_{\beta\alpha} \hat{p}_\beta \right),
 \label{eq:temporal3}
\end{equation}
that is formally the same as the stationary condition imposed on  Eq.~(\ref{eq:basic1Gen}) for the static network case and is similar to Markov chain approaches used to solve contagion processes in multiplex and interconnected networks~\cite{Cozzo2013b,Wang2013,Granell2013}. Given that Eq.~(\ref{eq:temporal3}) formally describes a diffusion process on a static network of $NT$ nodes, we can then  follow~\cite{Wang2003,Gomez2010} and linearize Eq.~(\ref{eq:temporal3}) recovering the necessary and sufficient condition for the asymptotically stable disease-free solution, $\rho\left[\mathbf{M}\right]<1$~\cite{Elaydi}. This yields the threshold condition 
\begin{equation}
\rho[\mathbf{M}] = 1
 \label{eq:threshold}
\end{equation}
for the critical value of $\lambda$ above which the  transmission becomes epidemic~\cite{Wang2003,Gomez2010,Cozzo2013b,Wang2013,Granell2013}. Given the block structure of the matrix, it is possible to simplify the computation of the spectral radius of $\mathbf{M}$~\cite{Powell2011}:
\begin{equation}
 \rho\left[ \mathbf{M} \right] =  \rho\left[ \mathbf{P} \right]^{1/T}
 \label{eq:detM}
\end{equation}
where
\begin{equation}
\mathbf{P} = \prod_{t=1}^T \left( 1-\mu+\mathbf{\Lambda}_{t} \right).
\label{eq:infection_propagator}
\end{equation}
In the case of unweighted undirected network, $\mathbf{P}$ becomes $\mathbf{P} = \prod_{t=1}^T \left( 1-\mu+\lambda \mathbf{A}_{t} \right)$~\cite{Valdano2015}. This matrix has an important physical interpretation. Let us consider a time-respecting path from $i$ to $j$, lasting $T$ time steps and containing $a$ jumps and $T-a$ waiting times. We associate to this path the weight $\lambda^a (1-\mu)^{T-a}$,  representing the probability that the infection propagates along that path, from $i$ infectious at time $t=1$  to $j$ infectious at time $t=T$. The entry $P_{ij}$ is then the sum of all the time-respecting paths going from $(i,t=1)$ to $(j,t=T)$, each weighted as described. Therefore, it represents the total probability of $j$ being infectious at time $t=T$, given that the infection originated in $i$ infectious at time $t=1$. This is valid in the limit of small probabilities and non-interaction among paths.
$\mathbf{P}$ thus describes the infection propagation around the disease free state (i.e. $p\simeq0$) and within the quenched mean field framework where interactions among paths are disregarded. In light of this interpretation, we call $\mathbf{P}$ {\itshape infection propagator}. The accessibility matrix, defined in~\cite{Lentz2013}, is a particular case of infection propagator, when $\lambda=\mu=1$, i.e., when the spreading process is a deterministic exploration of the temporal network.
The generalization to weighted network is straightforward once the force of transmission on each link of Eq.~(\ref{eq:weight_binomial}) is taken into account.

\section{Infection propagator approach for SIRS and SIR dynamics}
\label{sec:SIR-SIRS}
Many pathogens leave recovered individuals immune to reinfection. Such immunity may last indefinitely, or for a limited amount of time and is modeled through the introduction of an additional compartment, the recovered (R) state~\cite{Keeling2008,Anderson1992}. Infectious nodes enter the recovered state with probability $\mu$,  becoming immune to re-infection. We also consider that they leave this state with probability $\omega$, returning to the susceptible state. Any value of $\omega>0$ describes a SIRS model, characterized by an average immunity period $1/\omega$. $\omega=0$ corresponds instead to the SIR model, where immunity is assumed to be permanent.
Markov chain equations for the SIRS model are as follows:
\begin{equation}
\begin{cases}
 p_{t,i} = \left(1-\mu\right)p_{t-1,i}+ \\
 \phantom{p_{t,i}}+ (1-p_{t-1,i}-q_{t-1,i}) \left[1-\prod_j \left(1- \Lambda_{t-1,ji} p_{t-1,j} \right)\right]; \\
 q_{t,i} = \mu p_{t-1,i} + (1-\omega) q_{t-1,i}.
\end{cases}
 \label{eq:temporal}
\end{equation}
In addition to $p_{t,i}$, we define $q_{i,t}$ as the probability of being in the recovered state at time $t$. The computation of the threshold is equivalent to the study of the stability of  the disease-free state $p_{t,i}=0$~\cite{Wang2003,Gomez2010}. Equations are therefore linearized around that point, making all  quadratic terms disappear. In the case of the SIRS model, the disease-free state is $p_{t,i}=q_{t,i}=0$ whose stability is now studied by linearizing in both $p_{t,i}$ and $q_{t,i}$. The linearized form of Eq.~(\ref{eq:temporal}) is the following:
\begin{equation}
\begin{cases}
 p_{t,i} \approx \sum_j \left(\Lambda_{t-1,ji}+(1-\mu)\delta_{ij}\right)p_{t-1,j} \\
 q_{t,i} = \mu p_{t-1,i} + (1-\omega) q_{t-1,i}.
\end{cases}
 \label{eq:temporal_lin}
\end{equation}
From this we see that the equation for $\mathbf{p}_t$ no longer contains $\mathbf{q}_t$, showing that the evolution of the number of infected around the disease-free state does not depend on the recovered individuals. This in turns implies that the recovered compartment does not impact the epidemic threshold, as in the static case~\cite{VanMieghem2014,Shu2014}.
As a result, the same infection propagator describing the SIS dynamics (Eq.~\ref{eq:infection_propagator}) can be used to compute the epidemic threshold of a SIRS compartmental model on a time-varying network. 

The SIR model can be considered as a limiting case of the SIRS dynamics ($\omega \rightarrow 0$). The infection propagator for the SIRS model does not depend on the probability of waning of immunity $\omega$, as it only contains expressions in terms of $\lambda$ and $\mu$. The threshold computed with the infection propagator approach for the SIRS therefore holds for any arbitrarily small $\omega$. As a result, we can safely perform the following limit:
\begin{equation}
\lambda_{critical}^{SIR} = \lim_{\omega\rightarrow 0}\lambda_{critical}^{SIRS} = \lambda_{critical}^{SIRS} = \lambda_{critical}^{SIS}.
\end{equation}
Both SIRS and SIR models thus have the same threshold as the SIS compartmental model, not being affected by the recovery compartment and the duration of the immunity period.

\section{Application to empirical and synthetic model data}

We test the validity and accuracy of our predictions by comparing them with the results of explicit microscopic stochastic numerical simulations of the SIR and SIRS processes. We consider two temporal network models and one empirical time-varying network. In the following subsections we describe the data and methods considered and the corresponding results.

\subsection{Empirical and synthetic  model data}

We test our approach on two network models: the \ntag{activity} and  \ntag{bursty} models.
\ntag{activity} is built from the activity-driven model proposed by Perra \textit{et al.} in~\cite{Perra2012}. Each node is given an activity potential, drawn from a heterogeneous distribution. At each time step, nodes become active with a probability equal to their potentials. Active nodes establish $m_{stub}$ (here $m_{stub}=2$) connections with other nodes picked at random, and all links are renewed at every snapshot. We generate networks with $N=1000$ nodes and $T=20$ time snapshots, as in~\cite{Valdano2015}. In addition, we explored size effects by considering networks with sizes ranging from $N=10^2$ nodes to $N=10^4$ nodes. Activity potentials are assigned through the relation $a=1-e^{-\eta x}$, ($\eta=10$) and $x\sim x^{-\gamma}$ and $x\in[\epsilon,1]$ ($\gamma=2.8$ and $\epsilon=3\cdot 10^{-2}$). The obtained networks are characterized by a temporally uncorrelated sequence of snapshots yielding an aggregated network with heterogenous topology.

\ntag{bursty} is obtained from the model  introduced by Rocha \textit{et al.} in~\cite{Rocha2013}. Here, the probability of a node becoming active at time step $t$ is sampled from the distribution $(t-t')^{-\alpha_1}e^{-\alpha_2 (t-t')}$, where $t'$ is the time that node was last active. We consider networks of size $N=500$ and described by $T=50$ time snapshots, generated with $\alpha_1 = 2$ and $\alpha_2 = 5\cdot 10^{-4}$ as in~\cite{Valdano2015}. The obtained networks  account for a heterogeneous activation pattern describing a sequence of homogeneous networks where the inter-contact time is power-law distributed

In addition to the synthetic models above, we consider an empirical time-evolving network constructed from records of face-to-face proximity interactions between individuals in a high school during one day,  collected by Salath\'{e} \textit{et al.}~\cite{Salathe2010} (\ntag{school}). This network comprises $N=787$ nodes and we consider here $T=42$ time snapshots, each one of $10$ minutes. In Sec.~\ref{sec:aggregation} we will also examine the impact of a finer aggregation time.

\subsection{Numerical simulations}

We numerically simulate the disease diffusion of a SIR and of a SIRS infection dynamics on the above described networks. Simulations assume all individuals to be susceptible at the initial time, and  are seeded with an infected node chosen at random on the network. At each time step, infectious nodes can transmit the disease with probability $\lambda$ to their susceptible neighbors and recover with probability $\mu$. Here we consider unweighted networks, for the sake of simplicity. Weighted networks will be addressed in the next section in the study of time aggregation of the evolving network. In the SIRS model, recovered nodes turn  susceptible with probability $\omega$. Results of the simulations are obtained after randomizing the initial seed and the  time step of the $T$ sequence chosen as  the initial time step, and they are obtained under the assumption of periodic boundary conditions for network evolution. 

\begin{figure}[!ht]
\begin{center}
\resizebox{0.7\textwidth}{!}{\includegraphics{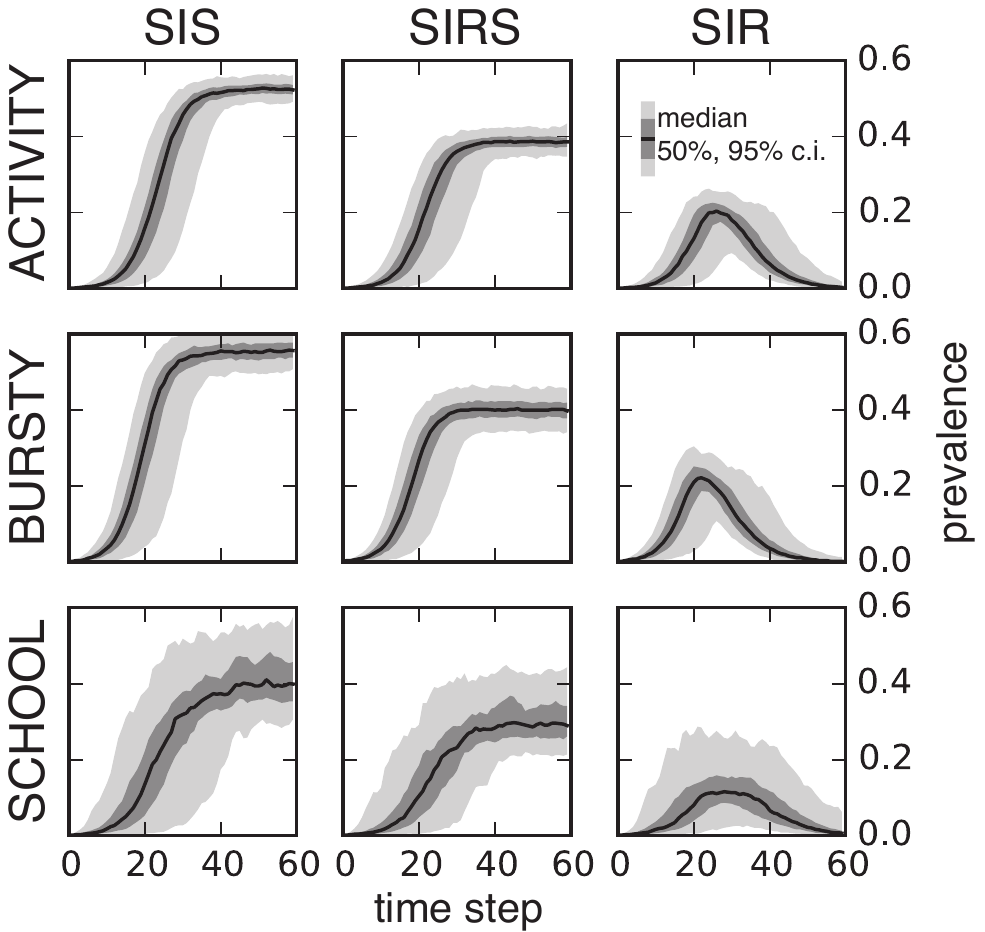}}
\end{center}
\caption{
{\bfseries Evolution of fraction of infected nodes (disease prevalence) for all considered disease models and networks.} In all panels $\mu=0.5$. For SIRS, $\omega=0.75$. $\lambda$ is set at $5/2$ of the epidemic threshold. This is an arbitrary value allowing us to show the typical behavior well above threshold. Median prevalence (solid black lines), and $50\%, 95\%$ confidence intervals (gray shaded areas) are computed on $200$ runs, keeping only the ones not going extinct in the beginning.
}
\label{fig:prevalence}
\end{figure}

Figure~\ref{fig:prevalence} shows the typical evolution of an outbreak, for the three disease models considered (SIS, SIRS, and SIR) and for all  networks. From this we clearly see that while above-threshold behavior is marked by a non-zero endemic state for both SIS and SIRS, a SIR outbreak shows the usual bell-like trend, reaching always extinction at the end.

For the SIRS dynamics, following~\cite{Ferreira2012} we numerically identify the epidemic threshold as the value of the transmissibility $\lambda$ for which the relative variation of the prevalence at equilibrium is maximal, as such variation would go to infinity in the thermodynamic limit ($N\rightarrow\infty$), indicating a second order phase transition. We therefore measure the {\itshape variablity}  $\Delta = \sqrt{\panino{i_T^2}-\panino{i_T}^2}/\panino{i_T}$~\cite{Ferreira2012,Crepey2006,Shu2014}, where $i_T = \frac{1}{T}\sum_{t=1}^T i_{eq}(t)$ is the prevalence at equilibrium averaged over a period $T$. The endemic prevalence is computed using the quasistationary method~\cite{Ferreira2011,Ferreira2012}. We force the system to be in an active state; whenever it reaches the absorbing state with no infectious nodes, we sample one random configuration among the ones the system had visited while it was in the same snapshot, and restart the simulation from that configuration. 
After discarding an initial transient ($3\cdot 10^3$ iterations), we compute $i_T$ for every period (for $5\cdot 10^5$ iterations), and with those values we compute $\panino{i_T}$ and $\panino{i_T^2}$. We then compare the value of $\lambda$ corresponding to the peak of the variability $\Delta$ with the prediction for the epidemic threshold obtained from the infection propagator approach. 
The same method is used for the SIR dynamics, where the endemic prevalence is replaced by the final attack rate $r$, i.e. the fraction of nodes hit by the epidemic. We stress that variability has formally the same definition in both SIRS and SIR models, but it is computed using different observables. We are only interested in the position of the peak~\cite{Ferreira2012}, as an indication of the epidemic threshold, and not its global behavior.

\subsection{Threshold results}

\begin{figure}[!ht]
\begin{center}
\resizebox{0.4\textwidth}{!}{\includegraphics{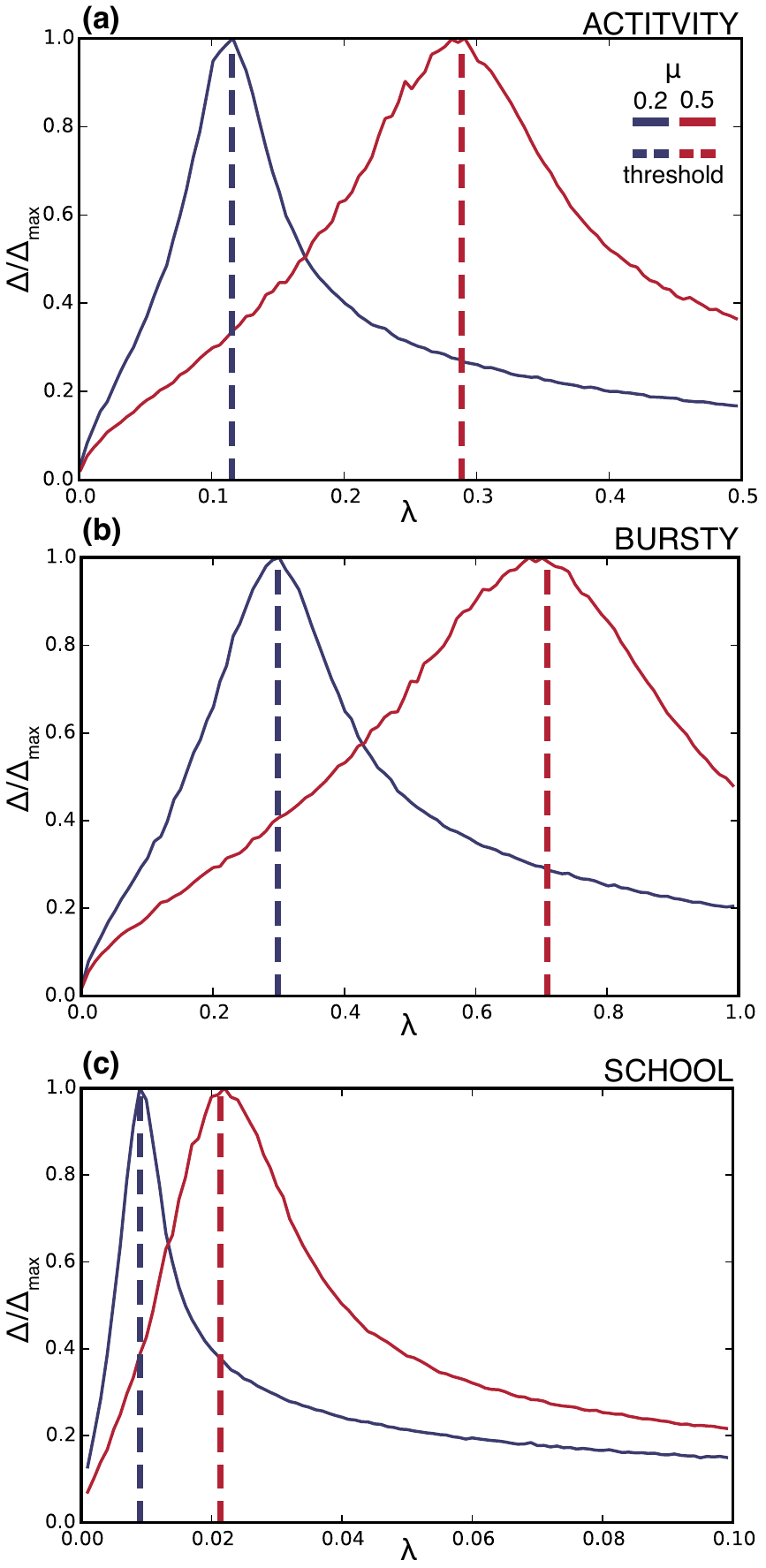}}
\end{center}
\caption{
{\bfseries SIR model. Comparison between the epidemic threshold and variability.} The variability $\Delta$ of final attack rate $r$, normalized to its peak value ($\Delta/\Delta_{max}$), is plotted against the transmissibility $\lambda$. We explore two different values of $\mu$. Dashed vertical lines represent the  threshold value for $\lambda$ predicted by the infection propagator approach. (a) shows the results for the \ntag{activity} network, (b) for the \ntag{bursty} network, and (c) for the \ntag{school} network.
}
\label{fig:th-SIR}
\end{figure}

We consider a SIR dynamics on the three networks under study and explore two values of the infectious period, corresponding to $\mu=0.2, 0.5$. For each $\mu$, Figure~\ref{fig:th-SIR} shows the behavior of the variability $\Delta$ normalized to its peak value $\Delta_{max}$ as a function of the transmissibility $\lambda$. In all cases we find a very good agreement between our prediction (vertical dashed line) and the simulated epidemic threshold obtained from the peak value of $\Delta$. The agreement is found for both network models, \ntag{activity} and \ntag{bursty}, despite them being characterized by different topological and temporal heterogeneities, and for the empirical dataset \ntag{school}. This last network features a more complex dynamics capturing the daily activities and interactions, with non-trivial temporal correlations and modular structures evolving in time~\cite{Salathe2010}. Despite the approximations used to compute the epidemic threshold with the infection propagator approach, the results of Figure~\ref{fig:th-SIR} indicate that the method is able to provide reliable and accurate predictions for the threshold behavior of systems characterized by different properties. We also note that the agreement is obtained independently of the values of the epidemic threshold: the threshold of the \ntag{school} network is indeed approximately one order of magnitude smaller than the ones obtained in the two network models for the same SIR dynamics. 

\begin{figure}[!ht]
\begin{center}
\resizebox{0.6\textwidth}{!}{\includegraphics{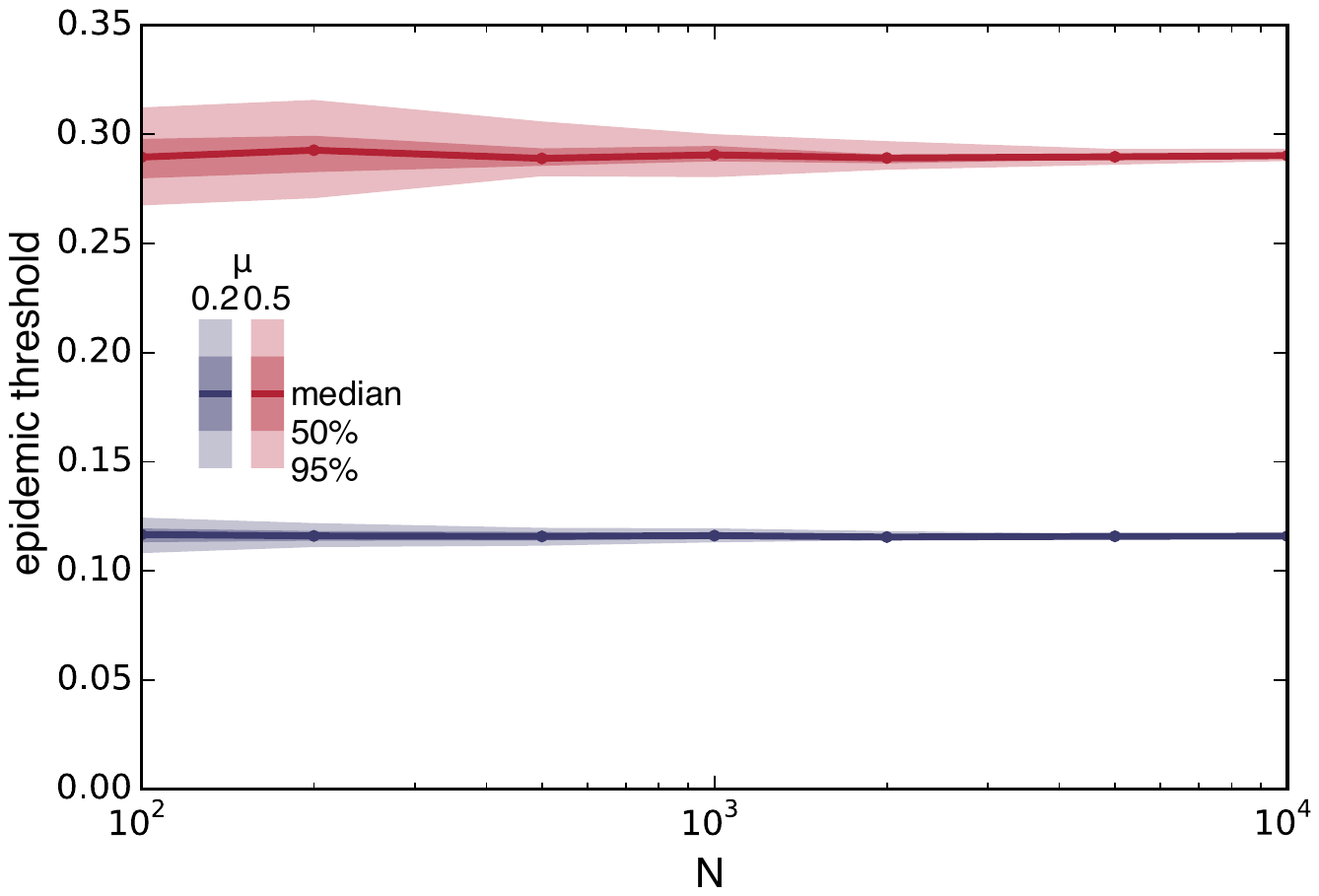}}
\end{center}
\caption{
{\bfseries Impact of the number of nodes on the epidemic threshold of the \ntag{activity} model.} 
We build \ntag{activity} models for different values of $N$, and compute their epidemic threshold for $\mu=0.2$. For each value of $N$ ($x$-axis) we build $200$ instances of the model, and compute their respective thresholds. On the $y$-axis we plot the epidemic the median, $50\%$ and $95\%$ of the threshold, computed over the instances.
}
\label{fig:explore_N}
\end{figure}

The choice of the number of nodes ($N$) and snapshots ($T$) in the models considered is  arbitrary, but it does not impact our findings. In~\cite{Valdano2015} we already studied the effect of $T$ on the threshold, and observed that, after an initial transient, the epidemic threshold saturates around a stable value, showing that the optimal $T$ has been reached. Here, in Fig.~\ref{fig:explore_N}, we explore the effect of $N$ for the \ntag{activity} model, by building different instances of the model, for several values of $N$. We find that the median value of the threshold, computed over the instances, does not depend on network size $N$. Fluctuations around the median, however, sharply decrease as $N$ increases, as expected.

\begin{figure}[!ht]
\begin{center}
\resizebox{0.4\textwidth}{!}{\includegraphics{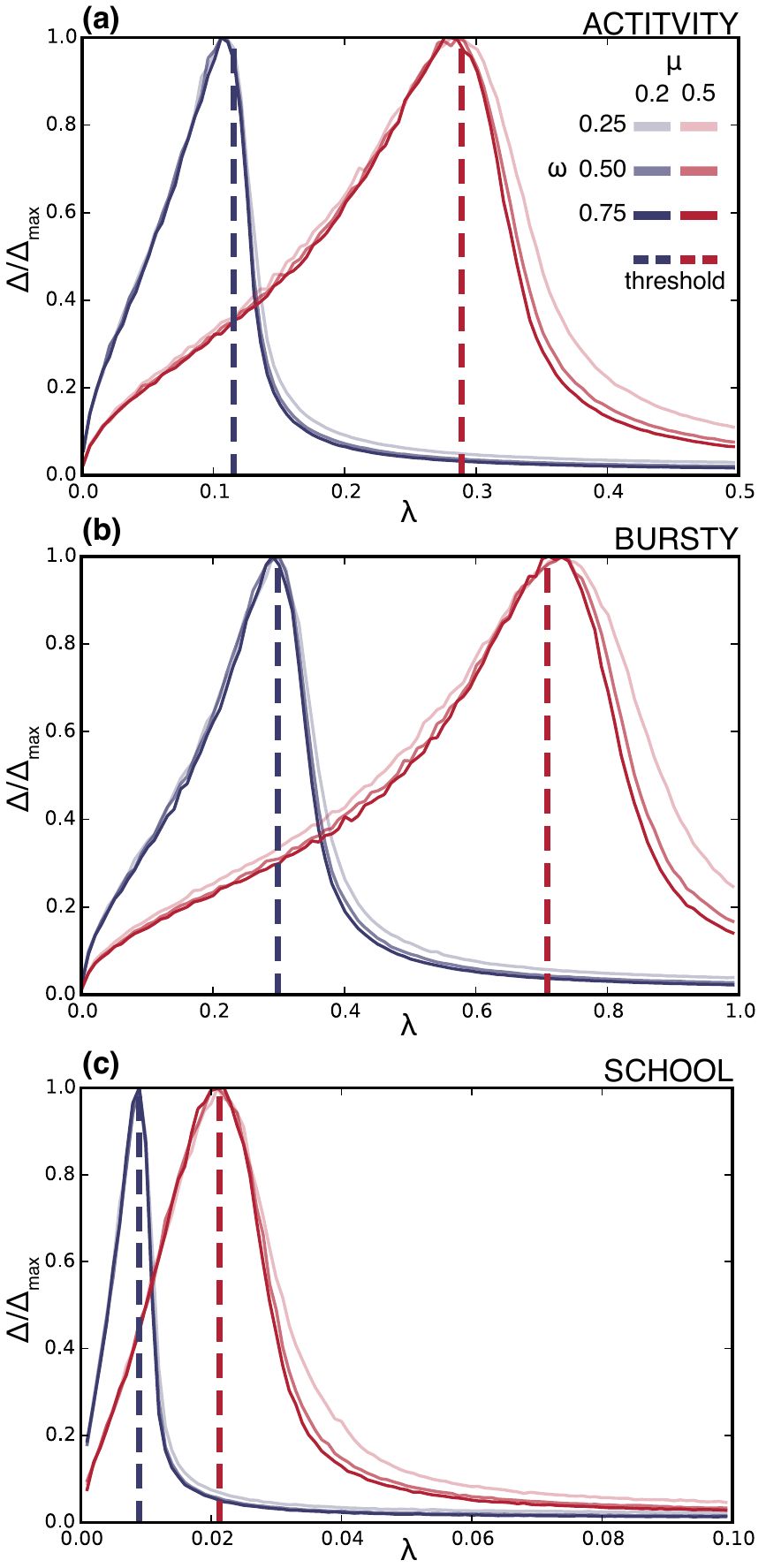}}
\end{center}
\caption{
{\bfseries SIRS model. Comparison between the epidemic threshold and variability.} The variability $\Delta$ of the endemic prevalence $i_T$, normalized to its peak value ($\Delta/\Delta_{max}$), is plotted against the transmissibility $\lambda$. We explore different values of $\mu$ and $\omega$. Dashed vertical lines represent the  threshold value for $\lambda$ predicted by the infection propagator approach. (a) shows the results for the \ntag{activity} network, (b) for the \ntag{bursty} network, and (c) for the \ntag{school} network.
}
\label{fig:th-SIRS}
\end{figure}

\begin{figure}[!ht]
\begin{center}
\resizebox{0.6\textwidth}{!}{\includegraphics{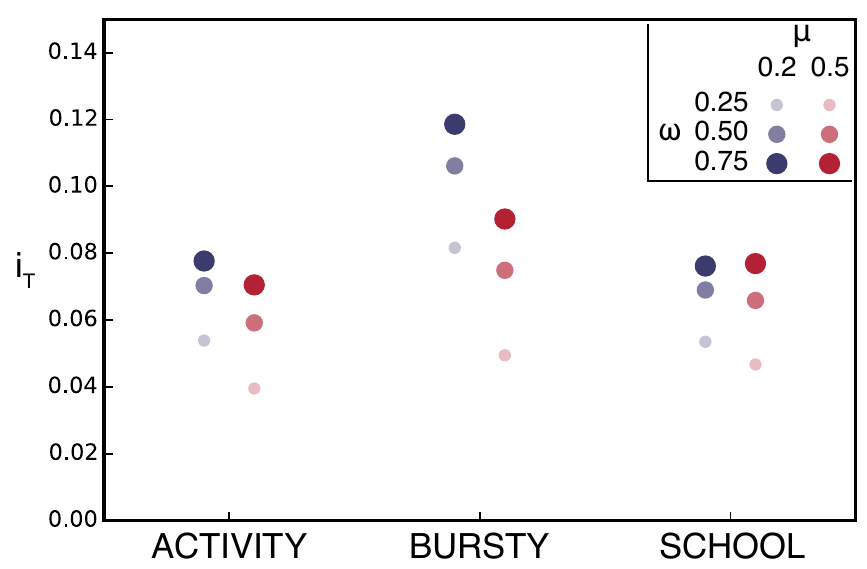}}
\end{center}
\caption{
{\bfseries SIRS model. Comparing the average prevalence above threshold, for different values of $\omega$.} For each network we choose a value of $\lambda$ above threshold, so that variability is around $1/3$ of its peak, $\Delta/\Delta_{max} \approx 1/3$. This value is clearly arbitrary, but it allows us to compare the different networks in similar epidemic situations. For \ntag{activity} this corresponds to $\lambda=0.14$ for $\mu=0.2$, and $\lambda=0.35$ for $\mu=0.5$; for \ntag{bursty}  $\lambda=0.36$ and $\lambda=0.86$, and for \ntag{school} $\lambda=0.01$ and $\lambda=0.03$. We plot the average prevalence at equilibrium $i_T$ for all three network, and for all the explored values of $\mu$ and $\omega$.
}
\label{fig:prevalence-SIRS}
\end{figure}

Results similar to those presented in Fig.~\ref{fig:th-SIR} are also obtained when considering a SIRS dynamics, characterized by the same values of the infectious period considered above and by three values of the probability of immunity waning ($\omega=0.25,\,0.5,\,0.75$). For each temporal network, we numerically identify the value of the epidemic threshold as that corresponding to the peak of the normalized variability $\Delta/\Delta_{max}$, and recover a good agreement with our analytical predictions (Figure~\ref{fig:th-SIRS}). The addition of the transition from an immune state to a susceptible state does not alter the accuracy of the computed predictions. Moreover, different immunity periods (i.e. different values of $\omega$) lead to the same epidemic threshold on the temporal networks, as predicted by the infection propagator approach. The  difference observed in the curves for different values of $\omega$ for $\lambda$ well above the threshold is induced by the variation in the average endemic prevalence. Epidemics circulating on these systems and characterized by longer immunity periods ($\omega=0.25$, light blue and light red in the plots of Figure~\ref{fig:th-SIRS}) display a larger variability due to the smaller average prevalence reached at equilibrium, as shown by Figure~\ref{fig:prevalence-SIRS}. Shorter immunity periods ($\omega=0.75$) reach a larger endemic prevalence for a given value of the transmissibility above the epidemic threshold and therefore display a smaller variability $\Delta/\Delta_{max}$. 

\section{Impact of time aggregation of the temporal network}
\label{sec:aggregation}

In many cases, information on network dynamics can be coarse, with data reporting on the temporal evolution at a lower resolution scale than the one of the process itself. This means that all events occurring within the time interval of the considered resolution will be aggregated in a static single snapshot. An aggregated representation of a temporal network does not account for causal structures and temporal correlations that occur at time scales that are smaller than aggregation interval~\cite{Holme2013}. Since these structures can impact disease dynamics, it is crucial to assess how such coarser representation influences the description of epidemic processes~\cite{Stehle2011,Ribeiro2013,Holme2013}. Here, we study the influence of the aggregation schemes described in~\cite{Stehle2011} on the epidemic threshold. \ntag{het} scheme is a weighted aggregation of the snapshots, obtained by summing link weights: $\mathbf{W}_{t}, \mathbf{W}_{t+1}\longmapsto \mathbf{W}_{t}+\mathbf{W}_{t+1}$. \ntag{hom} is topologically equivalent to \ntag{het}, having the exact same set of links. Each link is given an equal weight corresponding to the average link weight of the weight distribution of the \ntag{het} network aggregated over the same period. As a result, both schemes share the same average weight at every aggregation interval, but \ntag{het} accounts for weight heterogeneity. We use the intrinsic transmissibility $\lambda$ for comparison across different aggregation schemes and intervals, as it does not depend on weight. We consider the empirical dataset of the \ntag{school} network as it provides a richer temporal and topological set of features with respect to synthetic models. Also, the study on aggregation aims at providing useful practical information for data collection purposes. 

We consider the highest resolution network obtained from the \ntag{school} data corresponding to $\Delta t_1=20s$. Starting from this resolution, we aggregate snapshots recursively two by two, doubling the aggregation interval at each aggregation test. We consider the recovery rate $m$ as an intrinsic property of the disease,  thus  not changing with aggregation. The probability of recovery after a time $\Delta t$ is $me^{-m\Delta t}$. Aggregation interval at the $k$-th aggregation is $\Delta t_k = k\Delta t_1$. Hence, we compute the recovery probability at the $k$-th aggregation $\mu[\Delta t_k]$ as the probability of recovering within an interval $\Delta t_k$, i.e., $\mu[\Delta t_k] = 1-e^{-m\Delta t_1\, k}$. We explore four different recovery rates, $m = 1.8, 9, 18, 90$ h$^{-1}$, in order to explore different time scales of disease diffusion.
High recovery rates mean short average infectious periods, and thus fast disease progression at node level. Conversely, low recovery rates induce long infectious periods, resulting in slower microscopical disease dynamics. 

\subsection{Threshold results on aggregated networks}

\begin{figure}[!ht]
\begin{center}
\resizebox{0.4\textwidth}{!}{\includegraphics{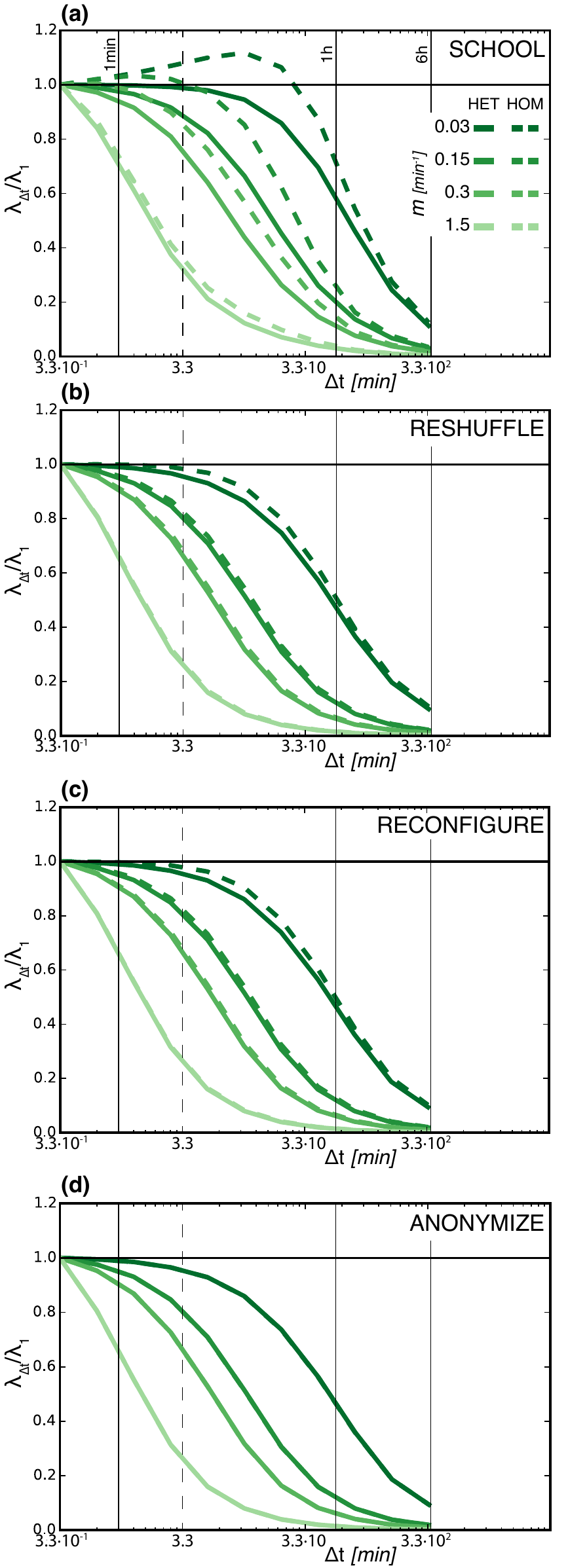}}
\end{center}
\caption{
{\bfseries Impact of aggregation on the epidemic threshold for the \ntag{school} network and three reference models.} The ratio between the threshold $\lambda_{\Delta t}$ computed on the aggregated network and the threshold $\lambda_1$ of the highest resolution temporal network is plotted as a function of the aggregation time interval $\Delta t$. Four different values of the recovery rates are explored, along with  two aggregation schemes, \ntag{hom} and \ntag{het}. (a) shows the results for \ntag{school}, (b) for reference model \ntag{reshuffle}, (c) for \ntag{reconfigure}, and (d) for \ntag{anonymize}. Black vertical solid lines mark aggregation intervals of $1min$, $1h$ and $6h$. Black vertical dashed line corresponds to $3min20s$.
}
\label{fig:SCHOOL-aggr}
\end{figure}

We compare the epidemic threshold $\lambda_{\Delta t}$ computed after aggregating the network with a given aggregation time window $\Delta t$, to the one computed at the highest resolution $\Delta t_1$ ($\lambda_1$), using the ratio $\lambda_{\Delta t}/\lambda_1$ (Figure~\ref{fig:SCHOOL-aggr}). 
For each recovery rate, the results from two time aggregation schemes are shown. 

Focusing on the \ntag{het} aggregation scheme, the results of Figure~\ref{fig:SCHOOL-aggr}(a) show that
the prediction made on the aggregated \ntag{school} network deteriorates with the increase of the time aggregation 
window $\Delta t$. As expected, the aggregation induces a loss of the temporal information 
making the aggregated network to perform poorly with respect to reproducing the behavior obtained in the
original network. This is known for a series of indicators regarding the importance of individual nodes in the
spread of an epidemic in the system~\cite{Holme2013}, and we find that it also results in a biased estimation 
of the  threshold condition for the epidemic propagation. 
The effect is more rapid and stronger for the
fast disease (e.g. $m=90$ hours$^{-1}$), in that it would have the possibility to experience the entire landscape of dynamical changes the network undergoes 
through, thus differentiating between the pattern obtained at the highest resolution and the aggregated one. 
On the other hand, for the slow disease (e.g. $m=1.8$ h$^{-1}$) we expect 
the epidemic process to be less sensitive to the network changes. The epidemic threshold computed 
on the aggregated network provides indeed a good estimate of the one corresponding to the highest
resolution network up to a certain level of aggregation (e.g., $\Delta t \simeq 3min20s$ for the $m=1.8$ hours$^{-1}$
case), after which the accuracy is progressively lost. This is consistent with the numerical results of an SEIR (susceptible--exposed--infected--recovered)
dynamics spreading on the network of contact of conference attendees, showing that the spreading dynamics
is well described by a static aggregated network if the heterogeneity of the contact durations is taken into account 
as edge weights~\cite{Stehle2011}.

The underlying mechanism leading to the deterioration of the epidemic threshold estimate with 
increasing aggregation time window is the creation of novel transmission paths that would otherwise
not exist, with the effect of destroying the causality of the sequence of interactions within $\Delta t$
and of increasing the density of the links in the network~\cite{Holme2005,Pan2011}. All these effects tend to facilitate the
spread of a disease, so that the resulting epidemic threshold is lower than the one computed on the
original temporal network corresponding to $\Delta t=\Delta t_1 = 20s$, as shown in Figure~\ref{fig:SCHOOL-aggr}(a).

If we focus on the \ntag{hom} aggregation scheme, we observe that the epidemic threshold predicted for a given
$\Delta t$ is systematically higher than the one obtained in the \ntag{het} scheme for the same $\Delta t$ value
(same color, dashed lines vs. continuous lines in Figure~\ref{fig:SCHOOL-aggr}(a)). The reason lies in the 
way  weights are distributed over the links of the aggregated networks. While the \ntag{het} scheme preserves the
heterogeneity of the duration of the contacts, cumulating the duration of the interaction established by each pair 
of individuals, this information is lost in the \ntag{hom} scheme as the total contact duration is homogeneously 
distributed among all contacts.
Heterogeneity of the weights has a strong effect on the evolution of epidemics~\cite{Mucha2010,Granovetter1973,Yan2005,Onnela2007,Lambiotte2009,Toivonen2009,Eames2009,Salathe2010b,Yang2012,Kamp2013} and in many cases it has the effect of favoring the spread of diseases~\cite{Ferreri2014,Deijfen2011,Britton2011,Britton2012}.
This results in a lower epidemic threshold than its homogeneous counterpart, for a given $\Delta t$.
The faster the disease is, the smaller is the difference observed in the epidemic threshold obtained from
the two aggregation schemes.

To better explore the various facets of the \ntag{school} temporal network having an impact on the threshold condition, we also consider three reference models that systematically destroy some of the network properties. \ntag{reshuffle} consists of a random reshuffling of snapshot time ordering. It preserves the aggregated network, and the static topological features of the snapshots. It breaks the temporal activity of the network, defined as the number of contacts in time. It breaks all temporal correlations among link activations, too. \ntag{reconfigure} consists of a random reassignment of contact timestamps. Two contacts $(i,j,t)$, $(k,l,s)$ are randomly selected, and their timestamp switched: $(i,j,s)$, $(k,l,t)$. It is equivalent to the {\itshape DCW} null model introduced in~\cite{Karsai2011}. \ntag{reconfigure} preserves the activity timeline and the aggregated network. It breaks snapshot topology and temporal correlations between link activations. Finally, \ntag{anonymize} reshuffles the identity of the nodes of each time snapshot, thus preserving activity timeline and static topology of each snapshot. It breaks all dynamic community structures and cliques (namely school classes). 

Results for these reference models are shown in Figure~\ref{fig:SCHOOL-aggr}(b)-(d). The behaviors obtained for \ntag{reshuffle} and \ntag{reconfigure} models are very similar. The difference between \ntag{het} and \ntag{hom} schemes is reduced for all recovery rates with respect to the results obtained on the original network, and it becomes negligible for faster diseases. The curves of panels (b) and (c) show that the obtained result is independent of the activity timeline of the network (preserved by the \ntag{reconfigure} reference model, but not by the \ntag{reshuffle} one), and it is more likely related to specific time-evolving topological structures present in the \ntag{school} network that are otherwise destroyed by both reference models. To test this hypothesis, we consider the \ntag{anonymize} reference model (Figure~\ref{fig:SCHOOL-aggr}(d)), where we destroy all two-points correlations and their time correlations, while preserving the overall temporal activity and the topology of each snapshot. As expected, the two schemes cannot be anymore distinguished following such reshuffling.

\begin{figure}[!ht]
\begin{center}
\resizebox{0.4\textwidth}{!}{\includegraphics{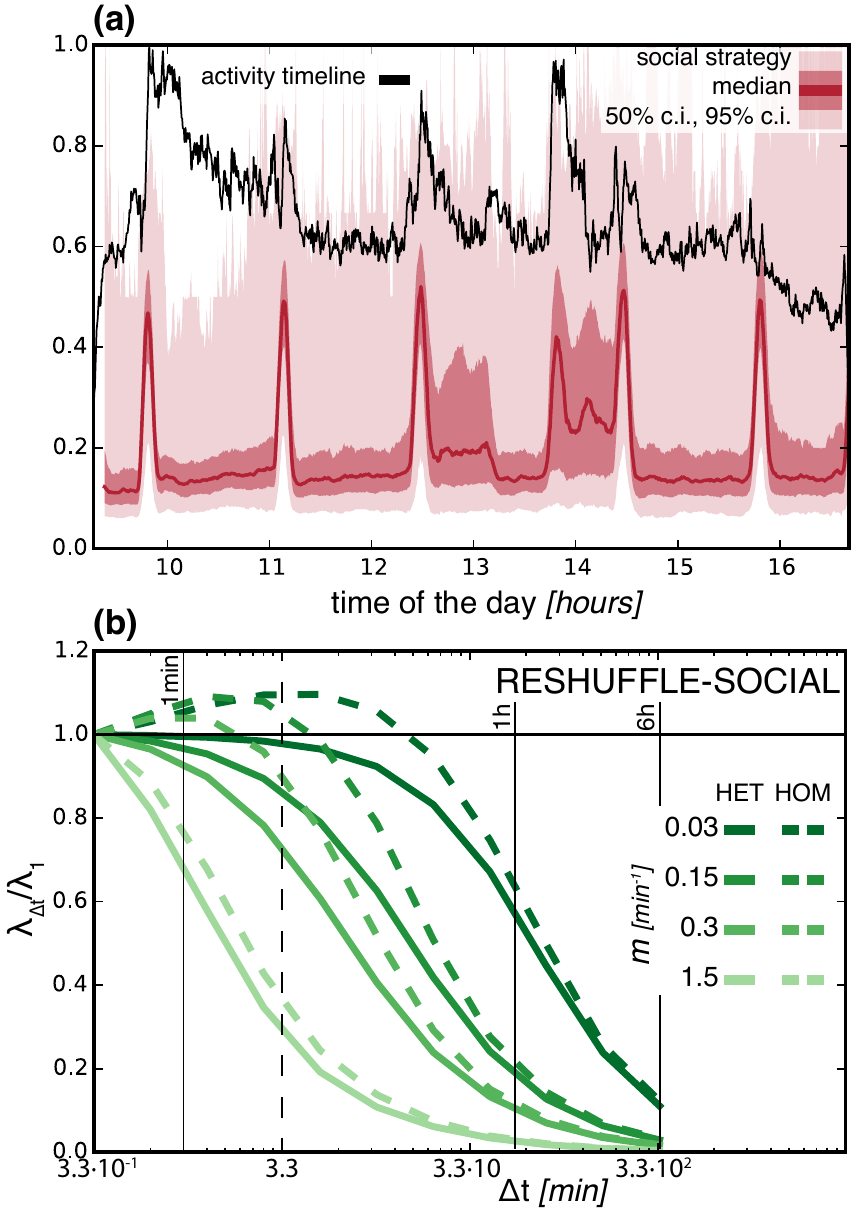}}
\end{center}
\caption{
{\bfseries Interplay between \ntag{school} dynamics and aggregation.} In (a) we show the temporal activity of \ntag{school}, as the normalized number of contacts (black line) in each snapshot of the fully temporal network. On the $x$ axis we indicate the time of the day, in hours. Red solid line represents the median value of network's social strategy, computed with a sliding window of 20 snapshots (equivalent to $400s$). Social strategy plotted at time $t$ is computed over the interval $(t-20,t]$. Red areas show $50\%$ (darker) and $95\%$ (lighter) confidence interval for social strategy. In (b) we plot, for reference model \ntag{reshuffle-social}, the ratio between the threshold computed on the aggregated network $\lambda_{\Delta t}$, and the threshold of the full temporal network $\lambda_{1}$, as a function of the aggregation time interval $\Delta t$. Four different values of the recovery rates are explored, along with the two aggregation schemes \ntag{hom} and \ntag{het}. Black vertical solid lines mark aggregation intervals of $1min$, $1h$ and $6h$. Black vertical dashed line corresponds to $3min20s$.
}
\label{fig:SCHOOL-LOC}
\end{figure}

Results of Figure~\ref{fig:SCHOOL-aggr} show that in all reference models aggregation leads to an underestimation of the epidemic threshold, for both aggregation schemes considered. In the \ntag{school} network, on the other hand, \ntag{hom} aggregation is found to provide larger epidemic thresholds than the one obtained at the highest resolution, within a given aggregation interval and for slow diseases. To better understand this behavior observed solely on the empirical data that disappears with the three types of reshuffling considered in the reference models, we explore the role of time correlations and memory effects in the \ntag{school} network. We consider the  {\itshape social strategy} introduced in~\cite{Miritello2013}. More in detail, we fix a time window of $\delta=20$ snapshots ($6min40s$), and define $k^{HOM}_{t,i}$ as the degree of node $i$ in the network aggregated over the interval $[t-\delta, t]$. The degree assumes the same value in both networks as it is the number of incident connections. We also define $s^{HET}_{t,i}$ as the weighted degree of node $i$ in \ntag{het} network, i.e., the sum of the weights of its incident links~\cite{Barrat2008}. We compute the social strategy of node $i$ at time step $t$ as $\gamma_{t,i} = k^{HOM}_{t,i}/s^{HET}_{t,i}$ (the same definition as in~\cite{Miritello2013}, except for a normalizing factor $\delta$). Social strategy discriminates between memory-driven behavior ($\gamma\rightarrow 0$), where a node tends to make contacts always with the same nodes, and  memoryless behavior ($\gamma\rightarrow 1$), where a node shows a more socially exploratory behavior. Figure~\ref{fig:SCHOOL-LOC}(a) shows how social strategy evolves in time. We observe that its median behavior is quite stable in time around low values, except for several localized spikes. Most of these spikes roughly correspond to abrupt variations in the temporal activity of the network. These spikes result then from a reduction of the memory of the system, due to a varying number of overall contacts. Remarkably, however, the median social strategy returns to the value it had before the spike  quickly after each of these events, indicating that the interaction dynamics does not qualitatively change, but  the sets of interacting individuals do change over time. The only exception occurs between around $13:30$ and $14:30$, when social strategy is significantly higher than average, but still lower than its delimiting peaks.
These spikes naturally induce a temporal slicing of the network, in a way that likely corresponds to the rhythm of school activities.
We call {\itshape $\gamma$-slice} each time interval between two consecutive spikes.

The degree of memory contained in the system, and measured through the tendency of each node to keep establishing contacts with the same individuals over time, is destroyed in all reference models under study, even those that preserve the activity timeline of Figure~\ref{fig:SCHOOL-LOC}(a). To understand whether if and to what extent stages in the time evolution of the social strategy  are responsible for the behavior observed in the \ntag{school} network, we design a fourth reference model, \ntag{reshuffle-social}, where we randomize the snapshot order, as in \ntag{reshuffle}, but we allow reshuffling only within each $\gamma$-slice.  Figure~\ref{fig:SCHOOL-LOC}(b) shows that \ntag{reshuffle-social} displays the same behavior as the \ntag{school} network, unlike \ntag{reshuffle}, with an overestimation of the value of the epidemic threshold by the \ntag{hom} scheme for small enough aggregation intervals and slow diseases. The aggregation of snapshots where individuals show a rather large memory in the way they establish links (i.e. small $\gamma$) leads to marked weight-topology correlations, likely being part of robust temporal communities of highly interacting nodes emerging from school daily activities. Such correlations were already found to play an important role in the slowing down of epidemics once large-scale propagation occurs in the system~\cite{Karsai2011}. In our case, we find that preserving the heterogeneity of weights of such correlations (as in the \ntag{het} scheme) can provide a good approximation of the epidemic threshold for small interval aggregation and for slow diseases. In addition, such approximation is better than the one provided by homogenizing  weights across all links in the system (as in the \ntag{hom} scheme), given that the latter destroys weight-topology correlations leading to a network that is more resilient to the epidemic spread~\cite{Ferreri2014}.
This effect vanishes for increasing time aggregating windows and it completely disappears for all aggregating intervals once these correlations are destroyed by the reshuffling of nodes (see the \ntag{anonymize} reference model in Figure~\ref{fig:SCHOOL-aggr}).

\section{Conclusions}

We have considered the infection propagator approach to compute the epidemic threshold for an arbitrary time-varying network. Starting from a SIS dynamics on a weighted directed temporal network, we have considered more complicated compartmental models and addressed timescale issues relevant for the study of temporal networks. The overall aim was to introduce the infection propagator approach for more realistic infection dynamics and to study the effect of time aggregation of the network of contacts on the computation of its threshold.  Our findings indicate that the approach provides reliable and accurate predictions of the epidemic threshold also in presence of immunity stages and loss of immunity transitions in the disease natural history. In addition, for slow diseases, the time aggregation scheme preserving the cumulative heterogeneous duration of contacts between two nodes is shown to provide a quite accurate estimation of the epidemic threshold of the corresponding high-resolution network up to a certain aggregation level. For faster diseases, time aggregation strongly alters the accuracy of the estimation. The presence of weight-topology correlations is the main feature of the \ntag{SCHOOL} network leading to biased estimations. These findings provide important information to study the vulnerability of systems in real settings and to assess possible biases induced by the consideration of time-aggregated contact data.

\newpage
\noindent
{\bfseries\sffamily Acknowledgements}\\
This work is partly sponsored by the EC-Health contract no. 278433 (PREDEMICS) to V.C.; the ANR contract no. ANR- 12-MONU-0018 (HARMSFLU) to V.C.; the EC-ANIHWA contract no. ANR-13-ANWA-0007-03 (LIVEepi) to  E.V., C.P., V.C.; the ``Pierre Louis'' School of Public Health of UPMC, Paris, France to E.V.

\newpage


\bibliography{valdano}{}
\bibliographystyle{unsrt}

\end{document}